# Of sequence and structure: Strategies of protein thermostability in evolutionary perspective

**Igor N. Berezovsky and Eugene I. Shakhnovich***

Department of Chemistry and Chemical Biology, Harvard University, 12 Oxford Street, Cambridge, MA 02138

*Correspondence should be directed to Prof. Eugene I. Shakhnovich, Department of Chemistry and Chemical Biology, Harvard University, 12 Oxford Street, Cambridge, MA 02138, phone: 617-495-4130, fax: 617-384-9228, e-mail:eugene@belok.harvard.edu

## Abstract

In this work we employ various methods of analysis (unfolding simulations and comparative analysis of structures and sequences of proteomes of thermophilic organisms) to show that organisms can follow two major strategies of thermophilic adaptation: (i) General, non-specific, structure-based, when proteomes of certain thermophilic organisms show significant structural bias toward proteins of higher compactness. In this case thermostability is achieved by greater overall number of stabilizing contacts, none of which may be especially strong, and (ii) Specific, sequence-based, whereby sequence variations aimed at strengthening specific types of interactions (e.g. electrostatics) are applied without significantly changing structures of proteins. The choice of a certain strategy is a direct consequence of evolutionary history and environmental conditions of particular (hyper) thermophilic species: ancient hyperthermophilic organisms that directly evolved in hot environment, pursued mostly structure-based strategy, while later evolved organisms whose thermophilic adaptation was a consequence of their recolonization of hot environment, pursued specific, sequence-based strategy of thermophilic adaptation.

## Introduction

The importance of various factors contributing to protein thermostability remains a subject of intense study (Elcock 1998; Jaenicke 1991; Jaenicke 1999; Jaenicke and Bohm 1998; Makhatadze and Privalov 1995; Szilagyi and Zavodszky 2000; Vogt and others 1997). The most frequently reported trends include increased van der Waals interactions (Berezovsky and others 1997), higher core hydrophobicity (Schumann and others 1993), additional networks of hydrogen bonds (Jaenicke 1999), enhanced secondary structure propensity (Querol and others 1996), ionic interactions (Vetriani and others 1998), increased packing density (Hurley and others 1992), and decreased length of surface loops (Thompson and Eisenberg 1999). Recently, it was demonstrated that proteins use various combinations of these mechanisms (England and others 2003a; Jaenicke 2000a; Vetriani and others 1998). However, no general physical mechanism for increasing thermostability (Jaenicke 2000a; Jaenicke 2000b) was found. The diversity of the "recipes" for thermostability immediately raises two important questions: (i) what is the common evolutionary or physical basis for the variety of mechanisms of thermostability, and (ii) how did this diversity appear and develop on the evolutionary scene?

To address the first question, one has to go beyond the analysis of specific stabilizing interactions and their various combinations. Conceptually, then, there can be two major factors that affect evolutionary selection of thermostable proteins. First, thermostable proteins may have structural bias such as enhanced packing. In this case, no single type of interaction may be extremely strong and dominate stabilization, but the sheer number of interactions provides enhanced stability. Second, stabilization can be achieved by very small number of particularly strong strategically placed interactions, e.g. electrostatics. This way, several substitutions made in sequences of mesophilic proteins can provide formation of “staples”, i.e. specific and strong interactions without significantly altering protein structure. We, therefore, posit two apparent possible scenarios for evolutionary selection of thermostable proteins: *structure-based (or non-specific) and sequence-based (or specific)*, each having their own advantages and drawbacks. Proteomes of thermostable organisms that were selected following first (structure-based) scenario would be enriched with proteins having enhanced structural features such as compactness. This mechanism of selection is *non-specific* in the sense that no or minimal distinct and special features of sequences are needed to achieve thermostability in sequence selection, making it robust under a wide range of environmental conditions. A possible evolutionary disadvantage of such a robust stabilization mechanism is that it makes proteins less adaptable to rapid and specific changes in environmental conditions. An alternative strategy may be sequence-based where structural repertoire of proteomes of thermostable organisms is not biased compared to their mesophilic counterparts. In this case, sequence selection plays major role whereby just a few strategic substitutions in sequence can lead to significant stabilization of an existing structure through the formation of several strong interactions *specific* to certain demands of the environment. These “staples” can work locally, leaving the bulk of the structure and its compactness unchanged.  There is, however, also a possible disadvantage to this mechanism. Sequence-based stabilization may not be robust because it is typically tailored to a specific and narrow range of environmental conditions.

The choice between specific, sequence-based, versus non-specific, structure-based, stabilization mechanism may be affected by a number of historical or environmental factors such as availability of the sequence/structure repertoire at different stages of protein evolution or a need to adapt to new environment (recolonization).

In this work we address the question of causal relationships between strategies of thermostability and their evolutionary context. (Shakhnovich and others 2004; Tiana and others 2004). By comparative analysis of sequences and structures of proteins from various (hyper) thermophilic organisms we indeed discovered two evolutionary strategies for achieving protein thermostability, structure-based and sequence-based, as outlined above. Further, we show how choice of a particular strategy for thermal adaptation can be understood in an evolutionary context.

## Materials and methods

### Simulations and sequence/structure analysis

The set of proteins we have analyzed in this work consists of 5 groups: 1. Hydrolase, from *E.coli* (1INO) and *T. thermophilus* (2PRD); 2. Rubredoxin, from *D. gigas* (1RDG), *C. pasteurianum* (5RXN), *D. vulgaris* (8RXN), and *P. furiosus* (1CAA); 3. 2Fe-2S Ferredoxin, from *S. platensis* (4FXC), *E. arvense* (1FRR), *Anabaena PCC7120* (1FRD), *H. marismortui* (1DOI), and *S. elongatus* (2CJN); 4. 4Fe-4S Ferredoxin, from *C. acidi-urici* (1FCA), *P. asaccharolyticus* (1DUR), *B. thermoproteolyticus* (1IQZ), *and T. maritima* (1VJW); 5. Chemotaxis protein, from *E. coli* (3CHY), *S. typhimurium* (2CHF), and *T. maritima* (1TMY). X-ray data from the Protein Data Bank were supplemented with coordinates of H-atoms (Berezovsky and others 1999).

Unfolding simulations were performed using an all-atom Gō model developed earlier (Shimada and others 2001). In the Gō interaction scheme atoms that are neighbors in the native structure are assumed to have attractive interactions. Hence Gō model of interactions is structure-based. Every unfolding run consists of $2x10^6$ steps. The move set contains one backbone move followed by one side-chain move. Van der Waals interactions were calculated for atoms belonging to residues separated by at least two residues along the polypeptide chain; only contact distances within 2.5-5.0 Å were considered for interactions (Berezovsky and others 1999).

High-throughput analyzes of the distributions of van der Waals contacts was performed on representative sets of major fold types, all α, all β, α/β, α+ β (according to SCOP classification (Murzin and others 1995), for list of the proteins used in the analysis see below), from *T. maritima*, *P. Furiosis/Horikoshii/Abyssi,* and *T. thermophilus*. Jack-knife tests have been performed to exclude: (i) possible effect of the same fold on the set, and (ii) influence of the size of the set.

Numbers of rotamers in fully unfolded states of Hydrolases (1INO and 2PRD) were calculated. Structures were unfolded at high temperature T=4 (see Figure 1a). Coordinate snapshots were recorded at every $10^5$ steps MC steps of total $10^7$ steps done for every structure. Numbers of rotamers for every residue were determined as an average over 100 snapshot.

Hydrogen bonds were determined according to criteria developed in (Berezovskii and others 1998; Stickle and others 1992).

Sequence alignments were done using software "MultAlign" developed in (Corpet 1988).

Distributions of number of van der Waals contacts per residue in archaea (from *P. furiosis/horikpshii/abissy*) *T. maritima* and *T. thermophilus* were calculated. Packing density (PD) is represented as number of contacts per residue. Number of contacts is normalized per PD bin (size of the bin is 30).

Designability has been treated within the frameworks of a residue-residue contact Hamiltonian (England and Shakhnovich 2003). It defines the conformational energy of a polypeptide chain to be the sum of the pair-wise interaction energies of all the amino acid pairs whose alpha carbins are separated by a distance less than ~7.5 Å (Miyazawa and Jernigan 1985).

**Listing of PDB-codes for major fold types in *P. abyssi/horikoshii/furiosis, T. maritima*, and *T. thermophilus***

Total numbers of analyzed folds from *P abyssi/horikoshii/furiosis*, *T. maritima*, and *T. thermophilus* are 37, 42, and 38, respectively. Numbers in the brackets show location of the fold in the structure. ***P. abyssi/horikoshii/furiosis* folds.** *All alpha:* 1AIS_B(1108-1300), 1AJ8, 1AOR(211-605), 1B43(220-339), 1I1G(2-61), 1IQP(89-169); *All beta:* 1B8A(1-103), 1DQ3_1(1-128), 1DQ3_2(415-454), 1DQI, 1ELT, 1H64, 1IQ8_1(506-582), 1IZ6(2-70), 1MXG(362-435), 1PLZ; *Alpha/beta:* 1A1S(1-150), 1A8L(1-119), 1E19, 1G2I, 1GDE, 1GEF, 1GTM(181-419), 1HG3, 1IM5, 1IOF, 1ION, 1IQ8_2, 1J08, 1JFL, 1JG1, 1LK5_1(1-130), 1LK5_2(211-229); *Alpha plus beta:* 1AIS_A(1-92), 1II7, 1K9X, 1NNW. ***T. maritima* folds.** *All alpha:* 1J5Y(3-67), 1JIO, 1M6Y_1(115-215), 1O0W(1-167), 1P2F_1(121-217), 1QC7; *All beta:* 1GJW_2(573-636), 1GUI, 1HH2(127-198), 1I8A, 1L1J, 1NCJ(2-101), 1O12_1(1-43), 1O12_2(332-364), 1O4T; *Alpha/beta:* 1A5Z(22-163), 1B9B, 1D1G, 1HDG, 1I4N, 1J9L, 1JCF(1-140), 1JG8, 1L9G, 1M6Y_2(2-114), 1M6Y_3(216-294), 1O0U, 1O14, 1O1X, 1O20, 1TMY, 1VPE; *Alpha plus beta:* 1DD5, 1GXJ, 1I58, 1J6R, 1M4Y, 1NZ0, 1O0X, 1O22, 1O26, 1VJW.
***T. thermophilus* folds.** *All alpha:* 1A8H(349-500), 1B7Y_B1(1-38), 1C52, 1DK1, 1EE8(122-210), 1GAX_1(797-862), 1GAX_2(579-796), 1IOM, 1IQR_1(172-416), 1IW7_E, 1N97, 1SES(1-110); *All beta:* 1EHK_B(41-168), 1EXM_1(213-312), 1EXM_2(313-405), 1FEU, 1GAX_3(190-342), 1IZ0_1(1-98), 1KWG(591-644), 1NYK, 2CUA, 2PRD; *Alpha/beta:* 1BXB, 1EXM_3(3-212), 1GAX_4(1-189), 1GAX_5(343-578), 1IQR_2(2-171), 1IR6, 1IUK, 1J09(1-305), 1J33, 1J3B, 1J3N, 1JL2, 1KA9_H, 1ODK, 1SRV, 1XAA.

## Results

The aims of our analysis were twofold: (i) to outline major strategies of protein thermostability, and (ii) to find an evolutionary basis for the development of particular strategies in the variety of species. These considerations defined the choice of the set of analyzed proteins. It includes five groups of proteins, each of them containing representatives of mesophilic organisms and its analogues from (hyper)thermophilic species. At the same time, members of these groups represent evolutionarily distant branches of the phylogenetic tree, archaea and bacteria.

### *Unfolding simulations with Gō model*

First, we evaluated stability of each of the proteins using an unfolding procedure based on the Gō model (Go and Abe 1981). According to the Gō model native interactions in the structure of the natural protein reflect mutually stabilizing effects of all or almost all types of interactions. It was demonstrated (Gō 1983) that Gō-like models that consider only native interactions give a satisfactory description of two-state folding processes of single-domain proteins. Thus Gō-model simulations aim at revealing *structure-based* contributions to protein stability, and, here, we started from the assumption that for the same reasons, it adequately reflects stability of the structure during its unfolding (Gō and Abe 1981).

Unfolding simulations for the studied groups of proteins reveal general trends of higher transition temperatures of unfolding for several (hyper)thermophilic proteins compared to their mesophilic counterparts. Figure 1a shows the difference between the hydrolases from thermophilic *T. thermophilus* and mesophilic *E.coli* towards higher stability of thermophilic protein. There is a pronounced difference between the unfolding temperatures of the rubredoxin from hyperthermophilic *P. furiosus* and rubredoxins from three mesophilic organisms (Figure 1b). Three mesophilic 2Fe-2S ferredoxins (4FXC, 1FRR, and 1FRD) demonstrate a narrow range of transition temperatures, whereas the thermophilic one (2CJN) from cyanobacterium *S. elongatus* has a substantially higher temperature of unfolding (Figure 1c). Analysis of 4Fe-4S ferredoxins from mesophilic and thermophilic organisms also reveals a significant difference in their transition temperatures (Figure 1d) demonstrating increased thermostability of thermophilic ferredoxin (1IQZ).

A striking exception from the general rule of higher simulation transition temperature for (hyper)thermostable proteins is represented by the proteins from hyperthermophilic *T. maritima*. Both 4Fe-4S ferredoxin (1VJW) and chemotaxis protein, CheY (1TMY), exhibit lower transition temperatures than their respective mesophilic counterparts (Figure 1d, e). Gō model discriminates, thus, proteins from *T. maritima* and demonstrates, that apparently mechanism of thermal stability for ferredoxin and CheY protein from *T. maritima* may be *different* from those of other (hyper)thermostable proteins studied in our unfolding simulations. Do proteins from *T.maritima* follow an alternative strategy to increase their thermostability? And if different strategies co-exist, what is the evolutionary basis for such different ways of thermal adaptation? First answers to these questions can be obtained from the analysis of the data presented in Table 1.

***Structural analysis***

According to the data in Table 1, hydrolase from the thermophilic bacteria has lower total van der Waals energy compared to its mesophilic counterpart. There are 6 $\alpha$–helices in thermophilic protein and only 3 $\alpha$-helices in the mesophilic one. Elements of secondary structure in thermostable hydrolase (2PRD) are rather extended in size, comprising 105 residues versus 84 in the case of the mesophilic protein (1INO). The total number of hydrogen bonds is also higher in a protein from the thermophilic organism: 170 versus 145. Thus, according to all structural factors presented in Table 1 hydrolase from *T. thermophilus* is expected to be more stable compared to its mesophilic counterpart. This also agrees with experimental data (Robic and others 2003) where role of the hydrophobic interaction in core region of thermophilic hydrolase was proven as a crucial factor of stabilization.

Another interesting feature of unfolding of hydrolases is almost complete coincidence of temperature-dependence curves of unfolding energies up to some relatively high temperature, followed by their abrupt separation. This can be explained by the difference in side-chain entropy of proteins due to the difference in their amino acid sequences. Calculation of average number of rotamers per residue in fully unfolded state (Canutescu and others 2003) gives values 12.0 and 11.4 for the mesophilic and the thermophilic proteins, respectively. It demonstrates, thus, higher side-chain entropy in the unfolded state of mesophilic hydrolase, which leads to its unfolding at lower temperature compared to thermophilic structure.

Hyperthermophilic rubredoxin from the archaebacteria *P. furiosus* demonstrates a pronounced bias towards high packing compared to mesophilic proteins (112 van der Waals contacts per residue in hyperthermophilic protein compared to 103, 98, and 96 in mesophilic analogues). Higher density of packing in hyperthermophilic proteins is also reflected in the increased number of H-bonds per residue and in the involvement of 62 per cent of residues into elements of secondary structure compared to 39-40 per cent in mesophilic proteins.

Van der Waals interactions and involvement of more residues into elements of secondary structure contribute to an increase of stability of thermophilic 2Fe-2S ferredoxin (2CJN, H-bonds can not be obtained because of low resolution NMR structure), in agreement with the conclusion done in experimental work (Hatanaka and others 1997).

All major structural factors presented in Table 1 point out to increased thermostability in thermophilic 4Fe-4S ferredoxin (1IQZ) and, thus, explain its higher transition temperatures in unfolding simulations compared to mesophilic analogues.

Proteins from *T. maritima* exhibit principally different distribution of major stabilizing interactions (Table 1). Analysis of the data for 4Fe-4S ferredoxin (1VJW) gives a substantially increased number of hydrogen bonds and involvement of almost half of the residues into secondary structure elements. At the same time, compactness of the structure (95 van der Waals contacts per residue in hyperthermophilic proteins compared to 96 and 82 in two mesophilic proteins) is practically the same as those in mesophilic protein. CheY protein (1TMY) has a decreased number of van der Waals contacts and hydrogen bonds, and slightly higher fraction of residues participating in secondary structure (see Table 1). Thus, both unfolding simulations (Figure 1) and structural analysis (Table 1) demonstrate that increased stability of thermophilic hydrolase (2PRD), ferredoxins (2CJN and 1IQZ), and hyperthermophilic rubredoxin (1CAA) from *P.*

*furiosis* is provided by the majority of structural factors acting together, whereas ferredoxin and CheY proteins from hyperthermophilic *T. maritima* lack structural connotation in their stabilizing mechanisms. This suggests that proteins from *T. maritima* have yet another way of increasing thermostability. In order to uncover a possible alternative mechanism of thermostability employed by *T.maritima* proteins we consider second major factor in protein stability, sequence.

***Sequence analysis***

We examined here sequence alignments of mesophilic proteins and their (hyper)thermophilic homologues (see Figure 2). Results of quantitative analysis of sequence comparisons are presented in Table 2. Similarly to unfolding simulations, sequence analysis discriminates proteins from hyperthermostable *T. maritima* from other (hyper)thermostable proteins analyzed in this work. Their sequences demonstrate pronounced difference in the alignments with their mesophilic counterparts (see the explanation of definition of residues in the Legend to Table 2). They have lower sequence identity with mesophilic proteins than other (hyper)thermophilic proteins (40 and 33 percent, percentage of residue types I and II in Table 2 summed up together, and positions colored by light and dark gray in Figure 2) for ferredoxin and CheY protein, respectively Moreover, 22 and 38 percent of sequence positions of *T. maritima* proteins do not match those in the sequences of mesophiles, while amino acids in the same positions of mesophilic sequences are identical to each other (light blue, Figure 2). In addition, we obtained substantial redistribution and increased number of charged residues in CheY protein and almost twice greater number of charged residues (11 versus 6, see also Table 2 and Figure 2) in ferredoxin, both from *T. maritima*, contrasted to their mesophilic counterparts. High level of sequence variation compared to mesophilic orthologs and significant bias towards charged residues in their sequences point out to key role of sequence selection in adaptation of T. *maritima* proteins to extreme conditions of the environment, in contrast to other (hyper)thermophilic organisms such as P. *furiosis* and T. *thermophilus* where structural bias is more pronounced. Remarkably, this finding is completely supported by experimental data where decisive role of surface ion interactions in hyperthermostability of proteins from *T. maritima* was demonstrated (Macedo-Ribeiro and others 1996; Usher and others 1998).

Among other proteins with increased stability analyzed in this work are thermophilic hydrolase (2PRD, from *T. thermophilus*), ferredoxins (2CJN and 1IQZ, from *S. elongatus* and *B. thermoproteolyticus*, respectively), and hyperthermophilic rubredoxin (1CAA, from *P. furiosus*). They exhibit high level of sequence identity (up to 80 percent) with their mesophilic orthologues (residue types I and II in Table 2; light and dark gray in Figure 2). Further, no significant substitutions into charged residues in sequences of respective (hyper)thermophiles (2PRD, 2CJN, and 1CAA) were observed (positions marked by blue and red (Figure 2) and residue types IV and V in Table 2, respectively). Several additional charged residues in thermophilic ferredoxin (1IQZ) can be explained by significantly larger size of the protein (81 residues versus 55 in mesophilic homologues). However, substantial elevation of packing density normalized by number of residues (27 percent more of contact per residue) and other structural factors (see Table 1) are apparent crucial contributors to increased stability, as it was detected by unfolding simulations (Figure 1d). Moreover, in the case of hyperthermophilic rubredoxin from *P.furiosis* (1CAA) and thermophilic ferredoxin from *S. elongatus* (2CJN) sequences of mesophiles contain in common parts of the alignments even more charged residues than their (hyper)thermophilic homologues (11 and 10 versus 4 and 3 per cent, respectively). Thus, all the approaches used in this work, structure-based unfolding simulations, analysis of structural features, and sequence alignments consistently distinguish proteins of *T. maritima* from the other (hyper)thermophilic proteins according to the differences in the ways of gaining thermostability. In the first case of thermophilic hydrolase (2PRD), ferredoxins (2CJN and 1IQZ), and hyperthermophilic rubredoxin (1CAA), we have a general trend of increasing of transition temperature obtained in unfolding simulations with a Gō model, essentially structure-based approach. We also found, for these proteins, that all stabilizing structural factors act concurrently, which points to compactness as the most probable cause for *structure-based* original mechanism of higher stability.

In the second case of proteins from *T. maritima*, we did not observe structural connotation for the mechanism of thermostability. At the same time, we revealed a strong sequence bias in proteins from *T. maritima*, which demonstrated preference for some of the stabilizing interactions and not others: a mechanism that we define as *sequence-based* strategy.

While the differences between mechanisms of thermostability demonstrated in this study for several proteins are suggestive, a fully conclusive evidence can be obtained only from massive comparison of proteins from different species.

***High-throughput analysis of major folds***

Our previous analysis suggested dominance of structure-based strategy in hyperthermophilic archaea *P. furiosis* and in thermophilic bacteria *T. thermophilus*, but not in hyperthermophilic bacteria *T. maritima.* This defined a choice of organisms for comparison of structural features of proteomes, namely packing densities. To this end, we compared distributions of number in proteins from *Pyrococcus* (archaea), *T. maritima*, and *T. thermophilius* (bacteria). We analyzed here structures of elementary domains (Murzin and others 1995). By examining domains instead of entire proteins we minimize possible artifact arising form surface effects.

We analyzed distributions of van der Waals interactions in representative sets of major fold types (all α, all β, α/β, and α+β, see Table 3 ) from *T. maritima, P. furiosis/horikpshii/abissy*, and *T. thermophilus*. Figure 3 shows that distribution of number of van der Waals contacts per residue in archaea folds (here, from *P. furiosis/horikpshii/abissy*) has most significant shift toward higher packing density (PD) compared to respective distributions for major folds from *T. maritima* and *T. thermophilus*. This observation is in full agreement with (i) increased contact density observed in several thermophilic proteomes (England and others 2003), and (ii) higher contact density for the last universal common ancestor (LUCA) domains /folds (Shakhnovich and others 2004). Remarkably, distribution of the number of contacts in the folds of thermophilic *T. thermophilus* is close to one for *Pyrococcus* folds, which indicates persistence of structure-based strategy in *T.thermophilus*. This finding is in full agreement with the conclusion obtained from unfolding simulations (Figure 1a) and structural analysis (Table 1). On the contrary, packing density in proteins from *T. maritima* is shifted toward lower values compared to both *Pyroccocus* and *T. thermophilus* folds. This observation suggests that proteins of hyperthermophilic *T.maritima* should apparently take alternative route to stabilization. The data presented in Table 3 clearly demonstrates quantitative difference in the distribution of number of contacts per residue. Proteins from *Pyrococcus* and *T. thermophilus* have higher mean values compared to proteins from *T.maritima* (275 and 272 contacts per residue versus 254, respectively). Besides, comparison of *T. maritima, Pyrococcus* and *T. thermophilus* proteins with those of mesophilic Yeast demonstrates that according to distributions of number of contacts *T. maritma* is closer to mesophilic organism by that parameter rather than to *Pyrococcus* and *T. thermophilus* (data not shown). Kolmogorov-Smirnov (KS) test shows high statistical significance of the difference between the distributions of contacts of the compared sets (see Table 3). This further proves persistence of structure-based strategy in *T. thermophilus,* whereas in *T. maritima* we found predominance of sequence-based mechanism.

***General concept of dual-strategy in thermostability***

The existence of the two mechanisms of thermophilic adaptation, structure-based and sequence-based, gives us an opportunity to look at adaptation process from the perspective of general concepts, structure and sequence. Using this approach, we can determine which strategy has been utilized by nature in any particular case, and how different strategies can be combined in order to reach adaptation to specific environmental conditions. As an example we take ferredoxins, whose universal presence in all organisms makes them an outstanding object for our analysis. There is a special interest in the group of 2Fe-2S ferredoxins, the ferredoxin from the halophilic archaebacterium *H. marismortui* (1DOI). First, this protein demonstrates a higher transition temperature (Figure 1) in unfolding simulations with structure-based Gō potential, which can be explained by significantly increased packing density and extensive hydrogen bonding (Table 1). It is worth noting that this halophilic protein is from archaebacteria, and it has substantially higher packing density than its mesophilic counterparts. This is another example (the first one is hyperthermophilic rubredoxin from archaebacteria *P. furiosus*) which corroborates the idea of high packing density as one of ancient mechanisms of thermostability (England and others 2003a). At the same time one can easily trace way of adaptation to high salinity. Almost entire surface of the protein is coated with acidic residues. This is achieved by enrichment of the sequence with acidic residues, in particular 8 of 22 residues in N-terminal domain are acidic, providing extra surface carboxylates for solvation. Thus, we observed co-existence of two stabilizing mechanisms: (i) specific, sequence-based, mainly by the abundance of acidic residues on the surface (Frolow and others 1996), which provides adaptation to high salinity, and (ii) non-specific, structure-based, which includes major factors of the protein stability and may well preserve stability and function of the protein under decreased salinity (Frolow and others 1996). This example highlights universality of two-strategy mechanism of adaptation, demonstrating versatility of adaptation to other than temperature factors of thermostability and provides a basis for its transformation into generic two-strategy mechanism of adaptation to wider spectrum of environmental conditions (temperature, salinity, pressure, etc.).

## Discussion

***Discriminative power of Gō model for variety of physical chemical factors of thermostability***

Earlier studies of the mechanisms of protein thermostability resulted in discovery of a variety of contributions to the effect (Berezovsky and others 1997; Elcock 1998; Hurley and others 1992; Jaenicke 1991; Jaenicke 1999; Jaenicke 2000a; Jaenicke 2000b; Jaenicke and Bohm 1998; Querol and others 1996; Schumann and others 1993; Szilagyi and Zavodszky 2000; Thompson and Eisenberg 1999; Vetriani and others 1998; Vogt and others 1997), and corresponding models on the basis of their combinations (Jaenicke 1991; Jaenicke and Bohm 1998). However, the diversity of protein folds of thermostable proteins, the mechanisms of stability, and evolutionary history of respective species raised questions about role of particular interactions or their combinations (Jaenicke 2000b). The elusiveness of universal rules of thermostability stems from the long-standing tendency to contrast the role of different stabilizing interactions, e.g. hydrophobic versus ionic interactions. Furthermore, an exceptional role in stabilization under high temperatures has been attributed exclusively to ionic interactions (Dominy and others 2004; Elcock 1998; Karshikoff and Ladenstein 2001; Perutz and Raidt 1975; Querol and others 1996; Xiao and Honig 1999; Zhou and Dong 2003). If that would be true, then one would have to universally observe prevalence of electrostatic stabilization in all thermostable proteins. However, in many of them this rule does not work (see Figure 1 and Table 1). High-throughput analysis on a proteomic level reinforces this observation (see Figure 3 and Table 3), showing apparent key role of increased packing density in achieving thermostability of proteins from hyperthermophilic archaea and thermophilic *T. thermophilus* in contrast to decrease of compactness coupled with prevalence of electrostatic interactions in *T. maritima.* This reveals, thus, an existence of several alternative ways of thermophilic adaptation. Here, we demonstrated how simple all-atom simulations can be used to estimate relative thermostability of the proteins in case of structure-based mechanism of stabilization. We considered here proteins from the species with different growth temperature: mesophilic (growth temperature up to 60°C), thermophilic (up to 80°C), and hyperthermophilic (more than 80°C). By analogy with microcalorimetric experiments (Privalov and Privalov 2000), where the transition temperature of unfolding is used as one of the parameters to evaluate protein thermostability, we compared transition temperatures of unfolding obtained in simulations on the basis of the Gō model (Go 1983). It should be noted, that Gō model is a simple structure-based approach and, thus, reflects mostly enthalpic contribution to the free energy correlated with compactness of the structure and opposing entropic factors arising from backbone and side-chain degrees of freedom. The model is neither supposed to predict transition temperature, nor to describe dependence of hydrophobic or electrostatic interactions on temperature. Our aim, here, was to point out to different strategies of thermostability, and we showed that Gō model is a proper tool to achieve that end. We demonstrated here, that more dense proteins (from *Pyrococcus*, *H. marismortui* (archea), and *B. thermoproteolyticus T.thermophilus* (bacteria)), that are stabilized by mostly hydrophobic interactions, unfold at higher temperatures in Gō simulations. In contrast we show that Gō simulations do not show increase of transtition temperature in proteins from *T.maritima*. This finding suggests that mechanism of stabilization in *T.maritima* is *different* from that in proteins with high packing density. Further, our analysis provides a new insight into physical mechanisms of thermostabilization showing two major strategies of increasing protein stability. We found structure-based stabilization for thermophilic hydrolase from *T. thermophilus*, 2Fe-2S ferredoxin from *S. elongatus,* and 4Fe-4S ferredoxin from *B. thermoproteolyticus* (packing density and other structural features are significant contributors), and hyperthermophilic rubredoxin from *P.furiosis*, which feature more compact folds so that all stabilizing interactions contribute to enhanced thermostability (see Table 1). The Gō model simulations also indicated a possibility of an alternative strategy of specific stabilization, where protein sequences are selected in such a way to enhance only one or few types of interactions in order to adapt to very specific extreme conditions. In this case, sequence variation, a mechanism that can introduce particular stabilizing interactions regardless of the detail of the original structure, gives rise to sequence-based specific strategy. Hyperthermophilic ferredoxin and chemotaxis protein from *T. maritima* exemplify this mechanism of stabilization. Here, the obvious bias towards specific interactions couples with lack of non-specific structure-based stabilization. These results are corroborated by the experimental data, revealing that hyperthermostable ferredoxin from *T.maritima* at 25 °C is "thermodynamically not more stable than an average mesophilic protein" (Pfeil and others 1997) and "conventional explanations for the structural basis of enhanced thermostability" do not work in case of chemotaxis protein from *T. maritima* (Usher and others 1998). At the same time, stability of these proteins under extremely high temperatures is provided by significant modifications of their sequences towards enrichment by charged residues, which turned out to be an effective sequence-based method of adaptation to extreme specific conditions (Pfeil and others 1997; Torrez and others 2003).

***Casual relationships between strategies of thermostability and their sequence/structure/evolutionary environments.***

What determines the choice of a strategy during long-time evolutionary experiment? Common believe that Life started from hot conditions implies two possible ways of evolutionary adaptation to hot environment: (i) first organisms whose adaptation mechanisms should be developed ''from scratch'', i.e. simultaneously with evolution of their proteomes, while (ii) on later stages organisms could recolonize extreme environment and, then, their already existing proteins should be changed. In the first scenario thermostable proteins were designed *de novo* – selection of sequence and structure had to occur concomitantly. This gives rise to evolutionary pressure on protein structures to make them more designable. Designability is a property of a protein structure that indicates how many sequences exist that fold into that structure at various levels of stability (Li and others 1996;  (Finkelstein and others 1995) England and Shakhnovich 2003; Taverna and Goldstein 2000).  Theoretical treatment of designability considers certain properties of contact matrix of a structure, C, (England and Shakhnovich 2003) as a major structural determinant of protein designability. Traces of powers of C reflect topological characteristics of the network of contacts within the structure, and, as a consequence, predict number of low-energy sequences that a fold can accommodate (England and Shakhnovich 2003). In particular, in lowest, second order in C approximation,  designability is predicted to correlate simply with compactness of a structure – number of contacts per residue (contact density) (England and others 2003b; Wolynes 1996). Figure 4 demonstrates that higher trace. i.e. more compact, structures (red diamonds) can obviously accommodate more low-energy sequences (gray shaded left part of the picture), than those of low contact trace. i.e. less compact structures (blue circles). This suggests that more designable structures were more amenable to become thermostable proteins at the early stages of evolutionary selection, when structures and sequences were selected concomitantly: more designable structures had initial advantage because greater number of sequences can fold into them with low energy, resulting in less severe sequence search requirements to make thermostable proteins having that structure. Together with earlier observation of higher contact density for last universal ancestor (LUCA) domains (Mirkin and others 2003; Shakhnovich and others 2004), it demonstrates that nature took advantage of higher designability in creation of first thermostable proteins of ancient species. Archaea proteins, rubredoxin from *P. furiosus* and 2Fe-2S ferredoxin from *H. marismortui*, exemplify this ancient mechanism of thermophilic adaptation, through selection of more compact (i.e. highly designable) structures (England and others 2003).

Second scenario is a modification of the existing proteins of an organism in response to abruptly changed conditions of the environment. The fast and effective way of tuning of protein stability without redesign of the whole structure is to make sequence substitutions which would lead to formation of “staples”, restricted set of specific interactions (*e.g.* ion bridges). This gives rise to sequence-based strategy of thermophilic adaptation. A good example of such strategy is *T. maritima* that recolonized hot environment (Nelson and others 1999). A whole-genome similarity comparison demonstrates (Nelson and others 1999), that *T. maritima* has *only* 24 per cent of genes that are most similar to Archaea’s. This similarity is a consequence of lateral (or horizontal) gene transfer (Lawrence and Ochman 1997; Nelson and others 1999), which, as it was demonstrated earlier, points to specific biochemical and environmental adaptations (Doolittle 1999a; Doolittle 1999b; Jain and others 1999; Lawrence 1999). In this case Archaea served as a source for lateral gene transfer on organismal level of adaptation during recolonization (Nelson and others 1999). However, mechanism of thermostabilization of remaning, biggest, part of its proteome should be developed, upon its colonization of hot environement, in *T. maritima* itself. In other words, when *T.maritima* recolonized hot environment, stability of already existing proteins must be significantly improved. We showed here a crucial role of sequence-based strategy thermostability in proteins from *T. maritima* versus structure-based one in Archaea proteins (see Results), which corroborates long evolutionary distance between T. maritima and Archaea (Nelson and others 1999).

Later in evolution structure-based strategy can persist in some cases, while it can be replaced by more specific, sequence-based, strategy in other cases (related to diverse environmental conditions and distinct evolutionary path they underwent). High-throughput structural analysis of major fold types implemented in this work provided the evidence of persistence/changing strategy of stabilization. We obtained non-specific structure-based mechanism in proteins of ancient Archaea (here, *Pyrococcus*) and its persistence and substantiation in bacteria *T. thermophilus.* At the same time this strategy was abandoned in other bacteria, *T. maritima,* where sequence-based strategy of implementing specific interactions was eventually developed. The latter represents, sophisticated mechanism of fine tuning of energetics and requires well-developed molecular mechanism of mutation/adaptation (Nelson and others 1999). Contrary to structure-based strategy, the key element here is a sequence variation that renders originally mesophilic protein a thermophilic one without significant alteration in its structure. A few specific interactions, as a result of sequence alteration, can crucially change stability of the structure, regardless of its original compactness and stability (Dominy and others 2004; Karshikoff and Ladenstein 2001; Macedo-Ribeiro and others 1996; Perutz and Raidt 1975; Usher and others 1998; Xiao and Honig 1999; Zhou and Dong 2003).

These findings and analysis highlights causal relationship between different strategies of thermophilic adaptation and evolutionary history of species.  Finally, coherent viewpoint into interplay of physical and evolutionary factors, provided by the two-strategy model, can be potentially helpful in guiding our effort to design proteins with desired thermal properties.

## Acknowledgements

The authors thank Jun Shimada for help with the unfolding simulations, Brian Dominy for the helpful discussions. Critical reading and valuable comments by William Chen and editing by Emmanuel Tannenbaum are greatly appreciated. This work is supported by NIH RO1 52126

## Figure legends

**Figure 1.** The temperature-dependence of the energy of unfolding. Every simulation of unfolding started from the native structure and included $2 \cdot 10^6$ MC steps. Absolute temperature increment is 0.2, and 0.1 in the vicinity of transition temperature. In all plots curves of the unfolding energy of mesophilic proteins are shown by black, blue, or cyan dots; thermophilic proteins – red dots; hyperthermophilic proteins – orange dots; halophilic protein – green dots. **(a)** Hydrolases, from *E.coli* (1INO, black rhombuses) and *T. thermophilus* (2PRD, red squares); **(b)** Rubredoxins, from *D. gigas* (1RDG, cyan triangulares), *C. pasteurianum* (5RXN, black rhombuses), *D. vulgaris* (8RXN, blue rhombuses), and *P. furiosus* (1CAA, orange squares). **(c)** 2Fe-2S Ferredoxin, from *S. platensis* (4FXC, cyan triangulares), *E. arvense* (1FRR, black rhombuses), *Anabaena PCC7120* (1FRD, blue rhombuses), *H. marismortui* (1DOI, green rhombuses), and *S. elongatus* (2CJN, red squares); **(d)** 4Fe-4S Ferredoxin, from *C. acidi-urici* (1FCA, black triangulares), *P. asaccharolyticus* (1DUR, blue rhjombuses), *B. thermoproteolyticus* (1IQZ, red squares), *and T. maritima* (1VJW, orange squares); **(e)** Chemotaxis protein, from *E. coli* (3CHY, blue rhombuses), *S. typhimurium* (2CHF, black squares), and *T. maritima* (1TMY, orange squares).

**Figure 2.** Sequence alignments for the groups of analyzed proteins. Only common parts of the alignments are presented and considered for calculation of Table 2. Letters are coloured as follows: light gray – the residue in the sequence of the (hyper)thermophile is identical to at least one of those in respective position of mesophilic sequences; dark gray – presence of charged residues in respective positions of (hyper)thermophilic and at least one of the mesophilic sequences; light blue – identical non-charged residues in respective

positions of at least two mesophilic sequences, while non-matching residue in (hyper)thermophile; blue – charged residue in at least one of the mesophiles, but non-charged residues in (hyper)thermophile; red – charged residues in the (hyper)thermophile, while non-charged residues in respective positions of mesophilic proteins. Bottom parts of the alignments contain information about secondary structure: dots show unstructured regions; letter E, residues involved into β-structure; H, elements of α-helices. **(a)** Hydrolases: 1ino(*Ec*) versus 2prd(*Tt*); **(b)** Rubredoxins: 1rdg(*Dg*)/5rxn(*Cp*)/8rxn(*Dv*) versus 1caa(*Pf*); **(c)** 4Fe-4S Ferredoxins: 1FCA(*Ca*)/1DUR(*Pa*) versus 1IQZ(*Bt*); **(d)** 2Fe-2S Ferredoxins: 1FRD(*Anabaena*)/4FXC(*Sp*)/1FRR(*Ea*) versus 2CJN (Se); **(e)** 2Fe-2S Ferredoxins 1FRD(*Anabaena*)/4FXC(*Sp*)/1FRR(*Ea*) versus 1DOI(*Hm*); **(f)** 4Fe-4S Ferredoxins: 1FCA(*Ca*)/1DUR(*Pa*) versus 1VJW(*Tm*); **(g)** Chemotaxis proteins: 3CHY(*Ec*)/2CHF(*St*) versus 1TMY(*Tm*).

**Figure 3.** Distribution of van der Waals contacts in representative sets of major fold types from *P. abyssi/horikoshii/furiosis* (red curve), *T. maritime* (black), and *T. thermophilus* (green). Packing density (PD) is represented through the number of contacts per residue. Number of residues is normalized per PD bin (size of the bin is 30).

**Figure 4.** Difference of sequence space entropy S(E) from its maximum value as a function of energy. Sequence space entropy S(E) represents logarithm of the number of sequences that can fold into a given structure with a given energy E. Red diamonds show S(E) for a more designable structure of high contact trace (or higher compactness in structural terms), blue circles correspond to structure of low contact trace. A greater number of low-energy sequences can be ’’accomodated’’ by higher trace structures (gray

shaded region), and, therefore, such structures can adopt a much larger number of foldable, highly thermostable sequences. The curves presented are for illustrative purposes only, detailed calculations for several specific models are presented in (England and Shakhnovich, 2003)

## Legends to Tables

**Table 1.** Factors possibly contributing of thermostability of analyzed proteins. Van der Waals interactions (Berezovsky and others 1999), number of H-bonds (Berezovskii and others 1998; Stickle and others 1992) and amount of residues involved into elements of secondary structure in groups of proteins under consideration. Parameters in the Table are as follows: vdW conts – total number of vdW contacts in protein; Cnts/res – number of vdW contacts per residue; N of bonds – number of H-bonds in protein; Bnds/res – number of H-bonds per residue; Sec. Strct – percentage of residues involved into the elements of secondary structure. Names of (hyper)thermpohilic organisms in column 2 are bolded italic. Numbers in brackets show difference between numbers of vdW contacts per residue, H-bonds per residue, and number of residues involved into secondary structurein mesophilic (averaged value was used if there are several mesophilic proteins in the group) and (hyper)thermophilic proteins, respectively.

**Table 2.** Quantitative results of the examination of sequence alignments for the groups of analyzed proteins (Column 1). Only common parts of the alignments (see Figure 2) are considered for calculation of Table 2. Types of residues are defined as follows: Type 1 (light gray in Figure 2 - residue in the sequence of the (hyper)thermophile is identical to at least one of those in respective position of mesophilic sequences; Type 2 (dark gray, Figure 2) – presence of charged residues in respective positions of (hyper)thermophilic and at least one of the mesophilic sequences; Type 3 (light blue, Figure 2) – identical non-charged residues in respective positions of at least two mesophilic sequences, while non-matching residue in (hyper)thermophile; Type 4 (blue, Figure 2) – charged residue in at least one of the mesophiles, but non-charged residues in (hyper)thermophile; Type 5 (red, Figure 2) – charged residues in the (hyper)thermophile, while non-charged residues in respective positions of mesophilic proteins.

**Table 3.** Comparative analyzes of the distributions of van der Waals contacts in representatives of the major fold types from *P. abyssi/horikoshii/furiosis, T. maritima*, and *T. thermophilus*. Kolmogorov-Smirnov test have been applied to united sets of the folds presented in each source. Results of the test are presented in a third column: number in brackets is P-value; *Tm* and *Tt* are *T. maritima* and *T. thermophilus*, respectively, and demonstrate differences between their proteins and those from the source (column 1). Last column (Fold types) demonstrates mean values of the distributions for major fold types in proteins from the respective organisms.

## Table 1

| Protein | Source | VdW energy | | Hydrogen bonds | | Sec. Strct |
|---|---|---|---|---|---|---|
| | | vdW conts | Cnts/res | N of bnds | Bnds /res | |
| **Hydrolase** | | | | | | |
| 1INO (175) | *E. coli* | 22804 | 130 | 145 | 0.83 | 0.48 |
| 2PRD(174) | ***T. thermophilus*** | 23178 | 133(2.3) | 170 | 0.98(18.1) | 0.6(25) |
| **Rubredoxin** | | | | | | |
| 1RDG (52) | *D. gigas* | 5363 | 103 | 40 | 0.77 | 0.40 |
| 5RXN (54) | *C. pasteuranium* | 5296 | 98 | 39 | 0.72 | 0.39 |
| 8RXN (55) | *D. vulagaris* | 5292 | 96 | 42 | 0.76 | 0.4 |
| 1CAA (53) | ***P. furiosus*** | 5914 | 112(13.1) | 45 | 0.85(13.3) | 0.62(56.3) |
| **Ferredoxin (2FE-2S)** | | | | | | |
| 4FXC (98) | *S. platensis* | 11005 | 113 | 76 | 0.78 | 0.37 |
| 1FRR (95) | *E. arvense* | 11767 | 124 | 96 | 1.01 | 0.43 |
| 1FRD (98) | *Anabaena PCC7120* | 12032 | 123 | 102 | 1.04 | 0.49 |
| 1DOI (128) | *H. marismortui* | 17537 | 137(14.2) | 131 | 1.02(8.1) | 0.5(16.3) |
| 2CJN (97) | ***S. elongatus*** | 13429 | 138(15.0) | - | - | 0.56(30.2) |
| **Ferredoxin (4FE-4S)** | | | | | | |
| 1FCA (55) | *C. acidiurici* | 5293 | 96 | 39 | 0.71 | 0.22 |
| 1DUR (55) | *P. asaccharolyticus* | 4507 | 82 | 37 | 0.67 | 0.4 |
| 1IQZ (81) | ***B. thermoproteolyticus*** | 9152 | 113(27.0) | 74 | 0.90(30.4) | 0.44(41.9) |
| 1VJW (59) | ***T. maritima*** | 5591 | 95(6.7) | 57 | 0.97(40.6) | 0.49(58.1) |
| **Chemotaxis Protein** | | | | | | |
| 3CHY (128) | *E. coli* | 17263 | 135 | 164 | 1.28 | 0.58 |
| 2CHF (128) | *S. typhimurium* | 17361 | 136 | 166 | 1.3 | 0.58 |
| 1TMY (118) | ***T. maritima*** | 15507 | 131(-3.3) | 134 | 1.14(-11.6) | 0.7(20.7) |

**Table 2**

| Proteins under comparison | Size of the common part of the alignments | Number of residues (percentage) | | | | |
|---|---|---|---|---|---|---|
| | | Type 1 | Type 2 | Type 3 | Type 4 | Type 5 |
| **1INO versus 2PRD** | 178 | 52 (29) | 38 (21) | - | 16 (9) | 16 (9) |
| **1RDG/5RXN/8RXN versus 1CAA** | 54 | 24 (44) | 17 (32) | 4 (7) | 6 (11) | 2 (4) |
| **1FCA/1DUR versus 1IQZ** | 58 | 15 (26) | 6 (10) | 15 (26) | 6 (10) | 10 (17) |
| **1FRD/4FXC/1FRR versus 2CJN** | 99 | 53 (54) | 26 (26) | 5 (5) | 10 (10) | 3 (3) |
| **1FRD/4FXC/1FRR versus 1DOI** | 99 | 30 (30) | 22 (22) | 22 (22) | 14 (14) | 9 (9) |
| **1FCA/1DUR versus 1VJW** | 60 | 18 (30) | 6 (10) | 13 (22) | 6 (10) | 11 (18) |
| **3CHY/2CHF versus 1TMY** | 124 | 24 (19) | 16 (14) | 47 (38) | 16 (14) | 19 (15) |

Table 3

| Source | Total number of folds or proteins /domains | Mean value of the distribution of number of vdW contacts per residue and KS-test (p-values) | Fold types | | | |
|---|---|---|---|---|---|---|
| | | | All α | All β | α/β | α+β |
| ***P. abyssi/furiosis/horikoshii*** | 37 | 275 *Tm* ($7.68 \cdot 10^{-2}$) *Tt* ($2.6 \cdot 10^{-2}$) | 282 | 238 | 284 | 296 |
| ***T. maritima*** | 42 | 254 *Tt* ($1.7 \cdot 10^{-1}$) | 269 | 236 | 265 | 233 |
| ***T. thermophilus*** | 38 | 272 | 273 | 269 | 276 | - |

**Figure 1.**

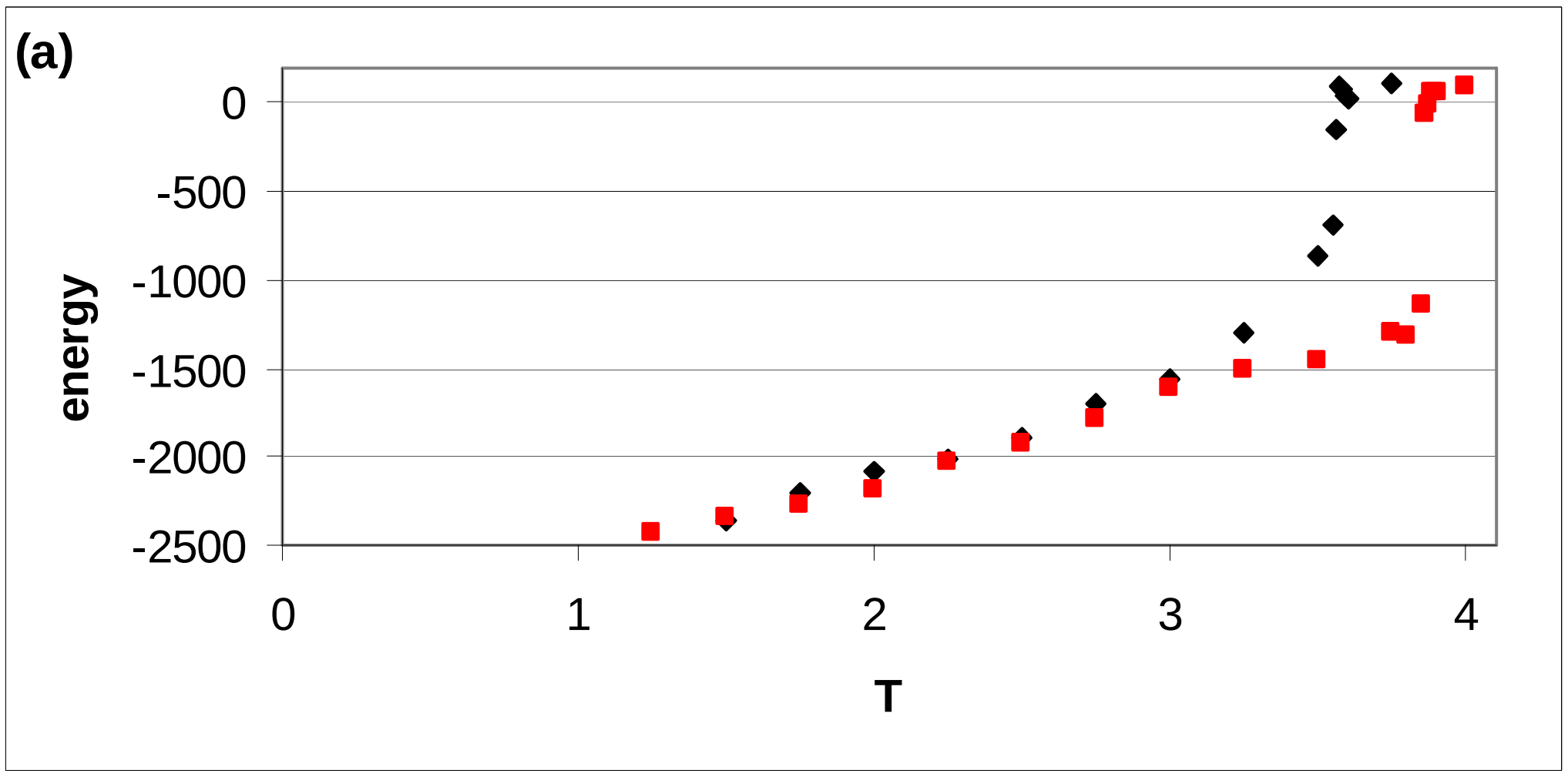

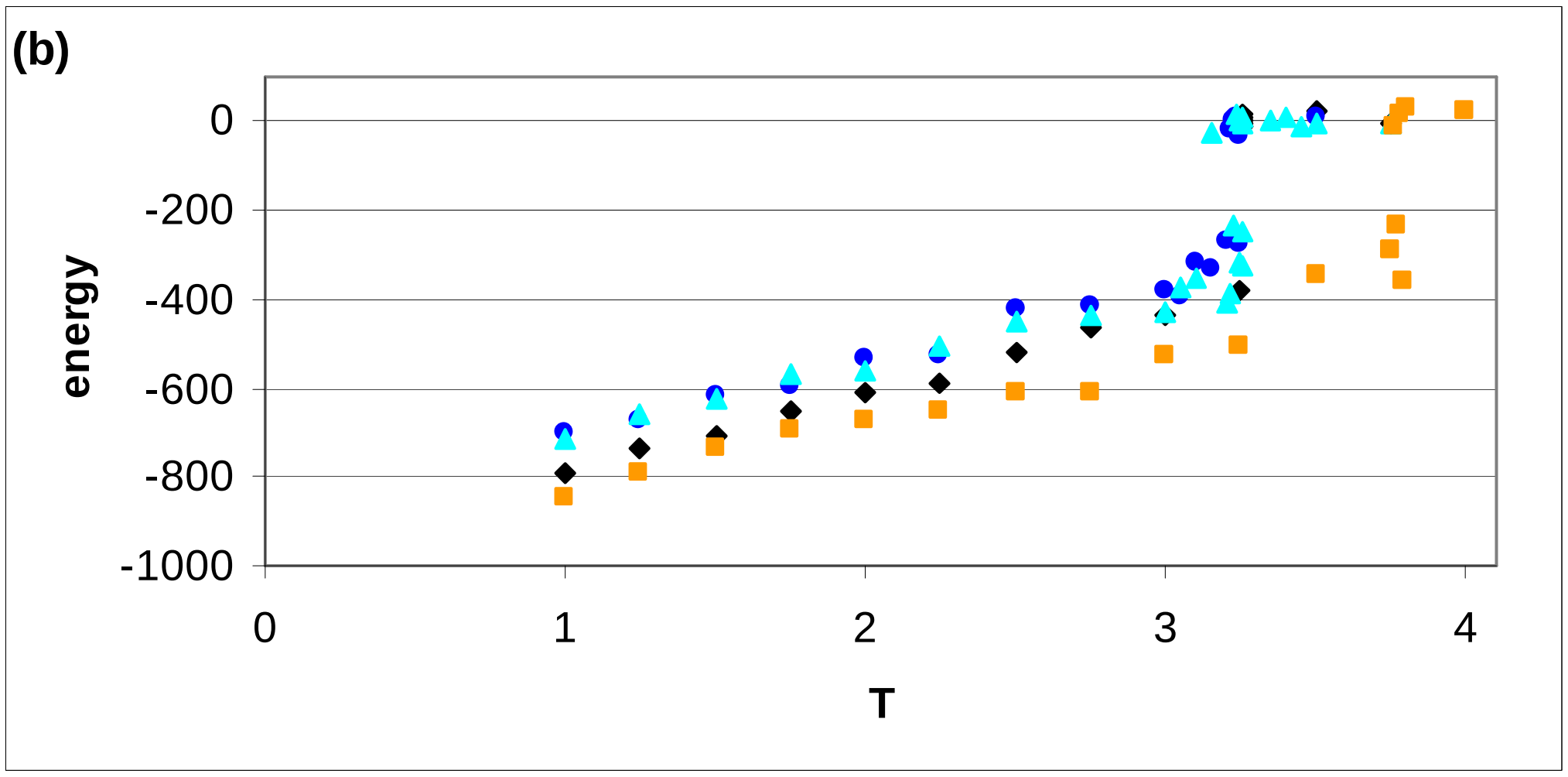

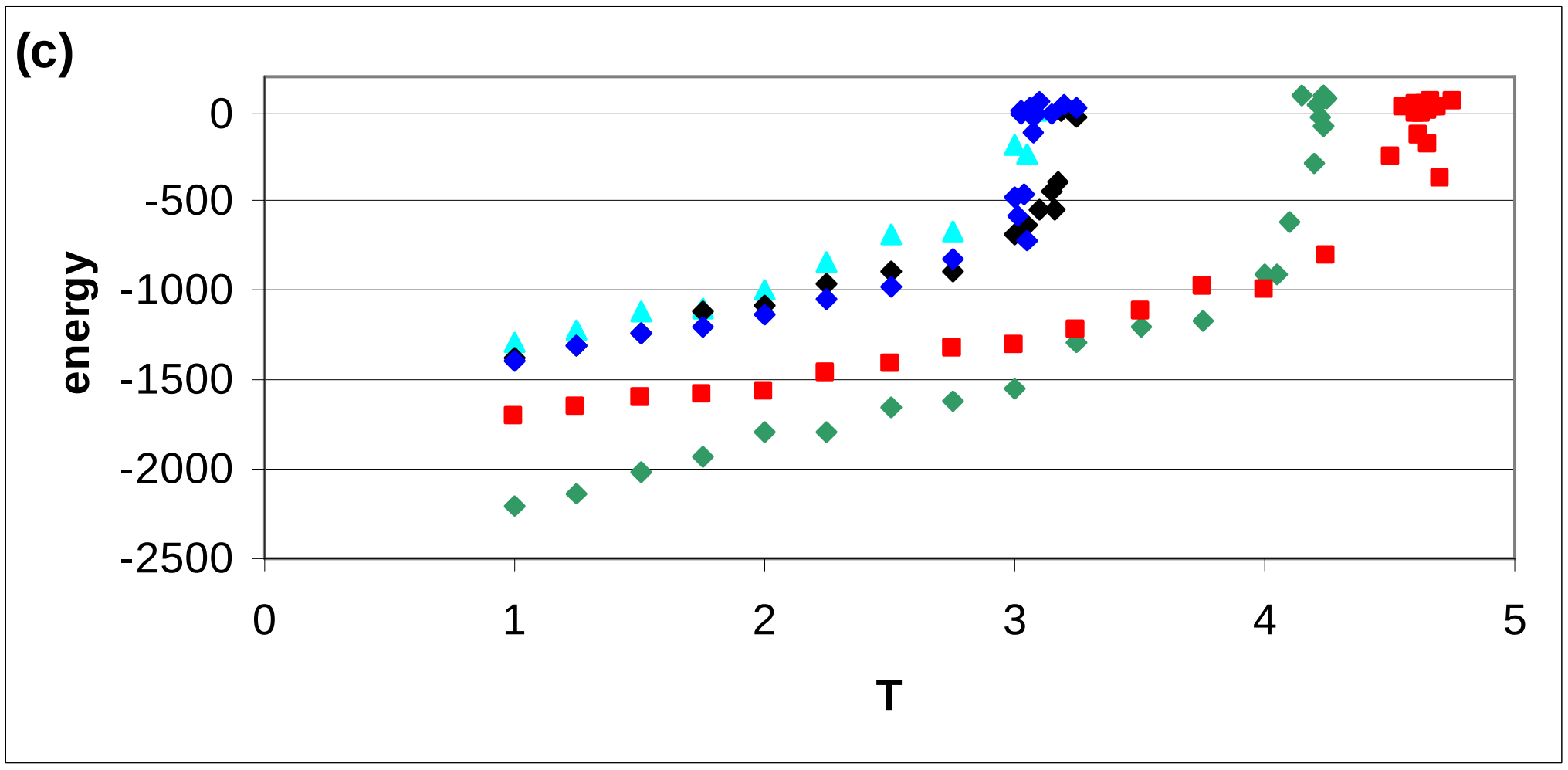

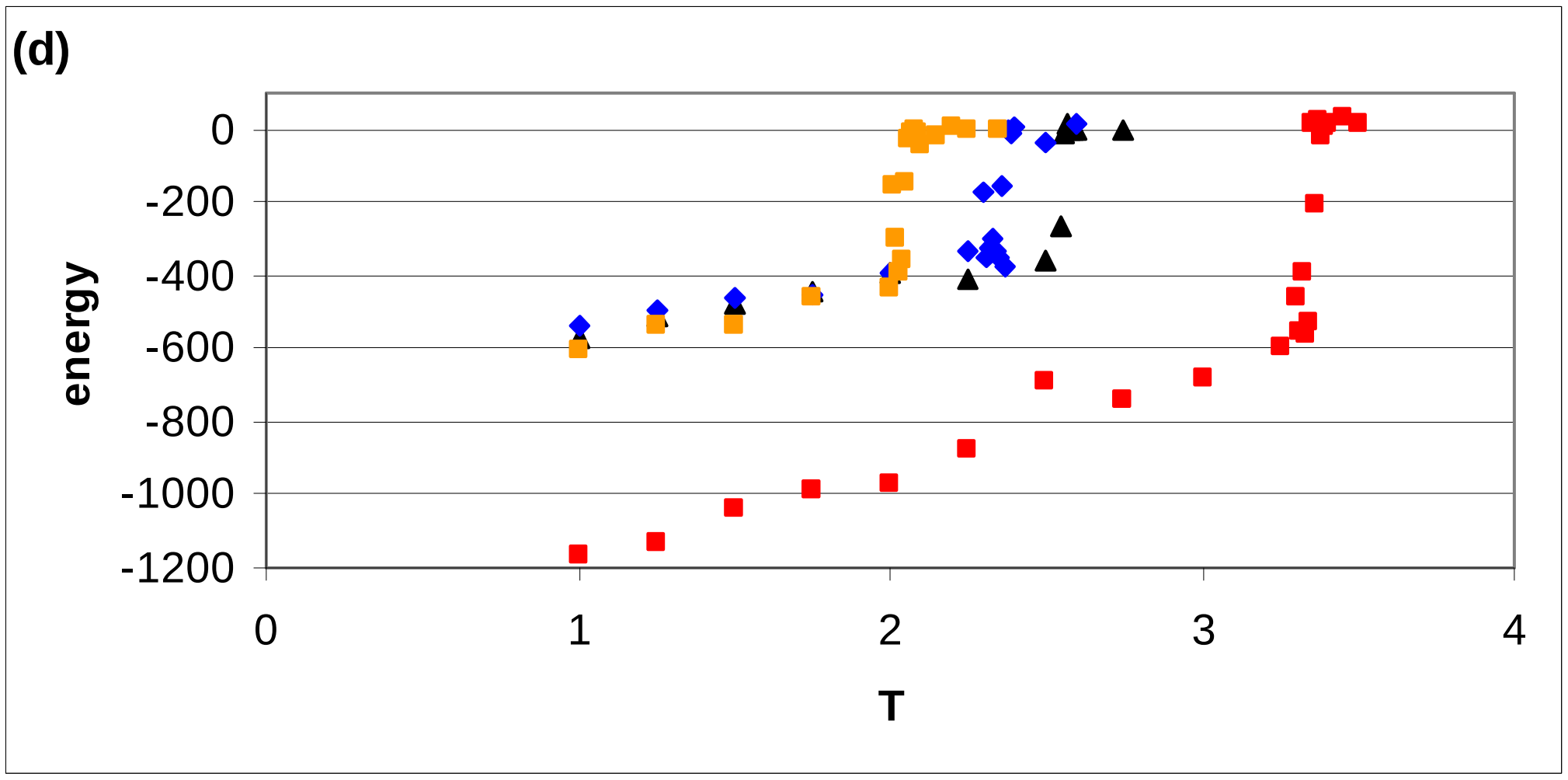

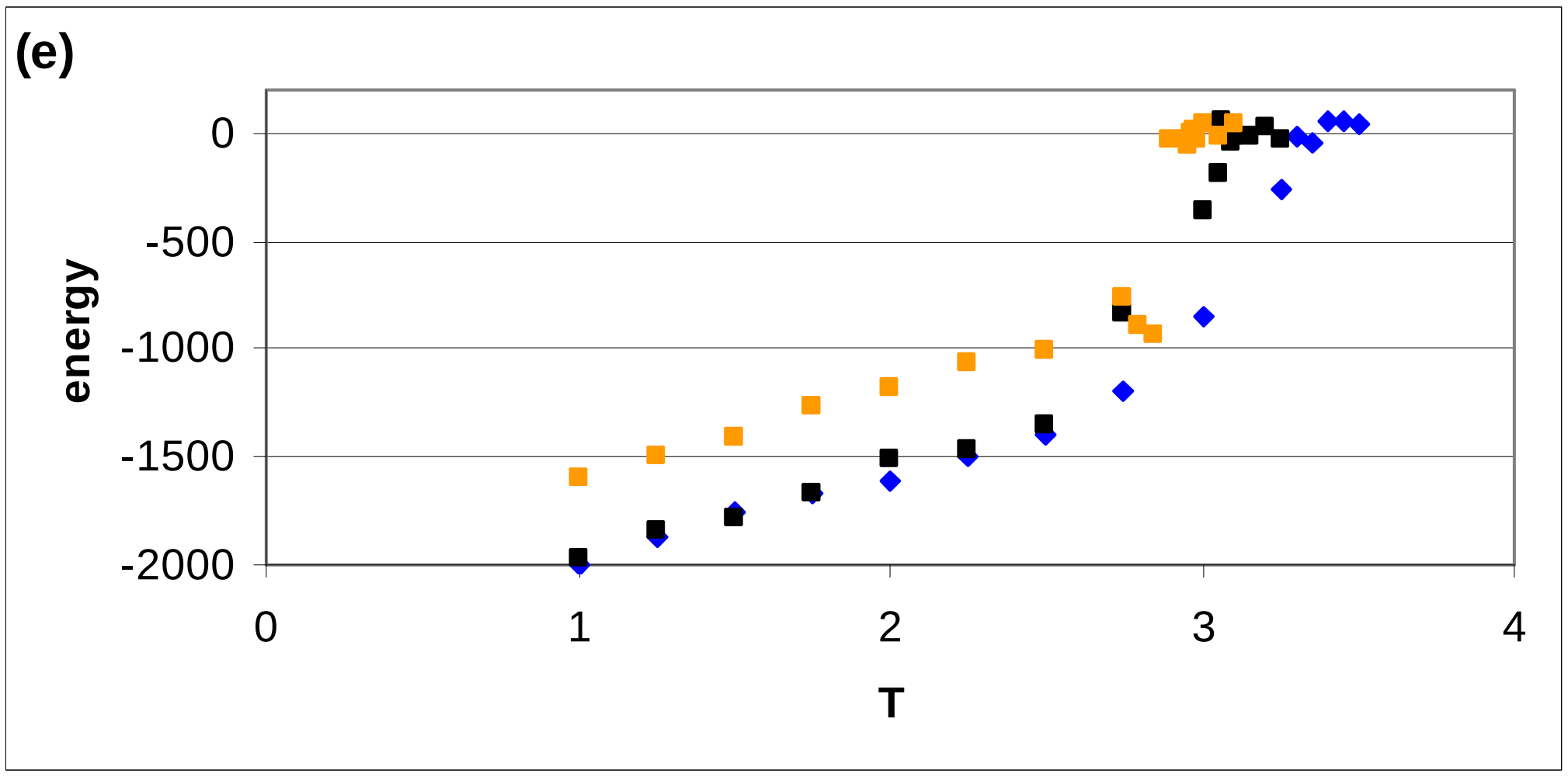

# Figure 2

**(a)**

```
1ino:  SLLNVPAGKDLPEDIYVVIEIPANADPIKYEIDKESGALFVDRFMSTAMFYPCNYGYINH
2prd: ANLKSLPVGDKAPEVVHMVIEVPRGSGN-KYEYDPDLGAIKLDRVLPGAQFYPGDYGFIPS
      =============================================================
1ino:  ..............EEEEEEE......EEEEE......EEEEEE.........EEEE...
2prd: ..HHH.........EEEEEEEE.....E-EEEEE....EEEEEEE.........EEEE...

1ino: TLSLDGDPVDVLVPTPYPLQPGSVIRCRPVGVLKMTDEAGEDAKLVAVPHSKLSKEYDHI
2prd: TLAEDGDPLDGLVLSTYPLLPGVVVEVRVVGLLLMEDEKGGDAKVIGVVAE--DQRLDHI
      ============================================================
1ino: .........EEEE..........EEEEEEEEEE.EEE......EEEEEE.....HHH...
2prd: .........EEEEE.........EEEEEEEEEEEEEE..EEEEEEEEEE..--.HHH...

1ino: KDVNDLPELLKAQIAHFFEHYKDLE--KGKWVKVEGWENAEAAKAEIVASFERAKNK
2prd: QDIGDVPEGVKQEIQHFFETYKALEAKKGKWVKVTGWRDRKAALEEVRACIARYKG
      =========================================================
1ino: .......HHHHHHHHHHHHH.....--....EEE..EEEHHHHHHHHHHHHHHHH..
2prd: ..HHH..HHHHHHHHHHHHH..HHHHHH...EEEEEEE.HHHHHHHHHHHHHHHH.
```

**(b)**

```
1rdg: MDIYVCTVCGYEYDPAKGDPDSGIKPGTKFEDLPDDWACPVCGASKDAFEKQ
5rxn: MKKYTCTVCGYIYDPEDGDPDDGVNPGTDFKDIPDDWVCPLCGVGKDEFEEVEE
8rxn: MKKYVCTVCGYEYDPAEGDPDNGVKPGTSFDDLPADWVCPVCGAPKSEFEAA
1caa:  AKWVCKICGYIYDEDAGDPDNGISPGTKFEELPDDWVCPICGAPKSEFEKLED
      ======================================================
1rdg: ...EEE.....EE.....EHHH.E.....HHH.....E......EHHHEEE.
5rxn: ...EEE.....EE.....EHHH.E.....HHH.....E......EHHHEEE...
8rxn: ...EEE.....EE.....EHHH.E.....HHH.....E......EHHHEEE.
1caa:  EEEEEE..EEEEEEHHHHHHHH.....HHHHH...........HHHHHEEE..
```

**(c)**

```
1fca: AYVINEACISCGACEPECP-VDAISQGGSRYVID---------ADTCIDCGACA-G
1dur: AYVINDSCIACGACKPECP-VNCI-QEGSIYAID---------ADSCIDCGSCA-S
1iqz: TIVDKETCIACGACGAAAPDIYDYDEDGIAYVTL*********PDILIDDMMDAFE
      ========================================================
1fca: .EEE..........HHH..-..............---------........HHH-H
1dur: .EEE.........HHHHH.-...E-E.....EE.---------.......HHHH-H
1iqz: EEE........HHHHHHH...EEE.....EEE..*********HHHHHHHHHHHHH

1fca: VCPVDAPVQA
1dur: VCPVGAPNPED
1iqz: GCPTDSIKVAD
      ===========
1fca: ......HHH.
1dur: HH....HHH..
1iqz: HH....HHH..
```

**(d)**

```
1frd: ASYQVRLINKKQDIDTTIEIDEETTILDGAEENGIELPFSCHSGSCSSCVGKVVEGEVDQ
4fxc: ATYKVTLINEAEGINETIDCDDDTYILDAAEEAGLDLPYSCRAGACSTCAGTITSGTIDQ
1frr:    AYKTVLKTPSGEFTLDVPEGTTILDAAEEAGYDLPFSCRAGACSSCLGKVVSGSVDE
2cjn: ATYKVTLVRP-DGSETTIDVPEDEYILDVAEEQGLDLPFSCRAGACSTCAGKLLEGEVDQ
      ============================================================
1frd: .EEEEEEEE....EEEEEEEE....HHHHHHH.................EEEE.E..EE.
4fxc: ..EEEEEEE....EEEEEEEE....HHHHHHH.................EEEE.......
1frr:    .EEEEEEE..EEEEEEE.....HHHHHHHH..................EEE......
2cjn: .EEEEEEEEE-..EEEEEEEEE...HHHHHHHH..................EEE..EEEE

1frd: SDQIFLDDEQMG-KGFALLCVTYPRSNCTIKTHQEPYLA
4fxc: SDQSFLDDDQIE-AGYVLTCVAYPTSDCTIKTHQEEGLY
1frr: SEGSFLDDGQME-EGFVLTCIAIPESDLVIETHKEEELF
2cjn: SDQSFLDDDQIE-KGFVLTCVAYPRSDCKILTNQEEELY
      =======================================
1frd: .......HHH..-..EEEHHH.EE...EEEE...HHH..
4fxc: .......HHHHH-.............EEEEE........
1frr: .......HHHHH-H............EEEEE...HHHHH
2cjn: .......HHHHH-H.......HHHHHHHHHHH...HHHH
```

**(e)**

```
1frd: ASYQVRLINKKQDIDTTIEIDEETTILDGAEENGIELPFSCHSGSCSSCVGKVVEGEVDQ
4fxc: ATYKVTLINEAEGINETIDCDDDTYILDAAEEAGLDLPYSCRAGACSTCAGTITSGTIDQ
1frr:    AYKTVLKTPSGEFTLDVPEGTTILDAAEEAGYDLPFSCRAGACSSCLGKVVSGSVDE
1doi: VFGEASDMDLDDEDYGSLEVNEGEYILEAAEAQGYDWPFSCRAGACANCAAIVLEGDIDM
      ============================================================
1frd: .EEEEEEEE....EEEEEEEE....HHHHHHH.................EEEE.E..EE.
4fxc: ..EEEEEEE....EEEEEEEE....HHHHHHH.................EEEE.......
1frr:    .EEEEEEE..EEEEEEE.....HHHHHHHH..................EEE......
1doi: HHHHHHH.......EEEEE......HHHHHHHH................EEEEEE..EEE

1frd: SDQIFLDDEQMG-KGFALLCVTYPRSNCTIKTHQEPYLA
4fxc: SDQSFLDDDQIE-AGYVLTCVAYPTSDCTIKTHQEEGLY
1frr: SEGSFLDDGQME-EGFVLTCIAIPESDLVIETHKEEELF
1doi: SDMQQILDEEVEDKNVRLTCIGSPDADEVKIVYNAKHL
      =======================================
1frd: .......HHH..-..EEEHHH.EE...EEEE...HHH..
4fxc: .......HHHHH-.............EEEEE........
1frr: .......HHHHH-H............EEEEE...HHHHH
1doi: .......HHHHH...EEE...EEE...EEEEEE.....H
```

**(f)**

```
1fca:   AYVINEACISCGACEPECP-VDAISQGGSRYVIDADTCIDCGA-CAGVCPVDAPVQA
1dur:   AYVINDSCIACGACKPECP-VNCI-QEGSIYAIDADSCIDCGS-CASVCPVGAPNPED
1vjw: MKVRVDADACIGCGVCENLCPDVFQLGDDGKAKVLQPETDLPCAKDAADSCPTGAISVEE
      ============================================================
1fca:    EEE..........HHH..-......................H-HHH......EEE.
1dur:    EEE.........HHHHH.-...E-E.....EE........HH-HHHHH....HHH..
1vjw: .EEEE.........HHHHH....EEEE....EEE.......HHHHHHHHH.....EEEE.
```

**(g)**

```
3chy: ADKELKFLVVDDFSTMRRIVRNLLKELGFNNVEEAEDGVDALNKLQAGGYGFVISDWNMP
2chf: ADKELKFLVVDDFSTMRRIVRNLLKELGFNNVEEAEDGVDALNKLQAGGFGFIISDWNMP
1tmy:   MGKRVLIVDDAAFMRMMLKDIITKAGYEVAGEATNGREAVEKYKELKPDIVTMDITMP
      ============================================================
3chy: ....EEEEE....HHHHHHHHHHHH.....EEEEE..HHHHHHHH......EEEEE....
2chf: ....EEEEE....HHHHHHHHHHHH.....EEEEE..HHHHHHHH......EEEEE....
1tmy:   ...EEEEE...HHHHHHHHHHHHH...EEEEEE..HHHHHHHHHHH...EEEEE...H

3chy: NMDGLELLKTIRADGAMSALPVLMVTAEAKKENIIAAAQAGASGYVVKPFTAATLEEKLN
2chf: NMDGLELLKTIRADSAMSALPVLMVTAEAKKENIIAAAQAGASGYVVKPFTAATLEEKLN
1tmy: EMNGIDAIKEIMKIDPNAK--IIVCSAMGQQAMVIEAIKAGAKDFIVKPFQPSRVVEALN
      ============================================================
3chy: ...HHHHHHHHH........EEEEEE....HHHHHHHH......EEE....HHHHHHHHH
2chf: ...HHHHHHHHH........EEEEEE....HHHHHHHH......EEE....HHHHHHHHH
1tmy: HH.HHHHHHHHHHH.....--EEEEE....HHHHHHHHHH...EEEE....HHHHHHHHH

3chy: KIFE
2chf: KIFE
1tmy: KVSK
      ====
3chy: HHHH
2chf: HHHH
1tmy: H...
```

**Figure 3**

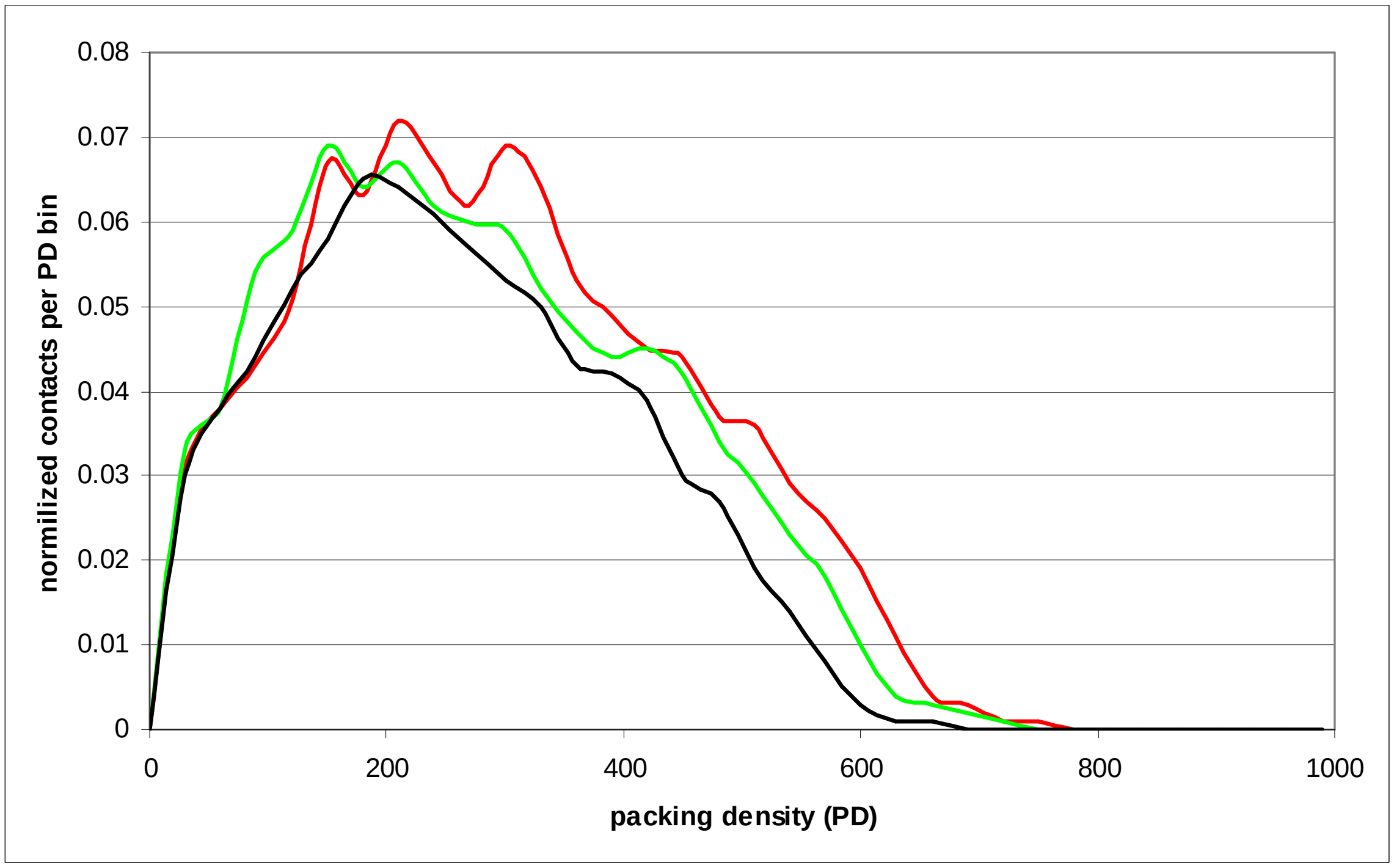

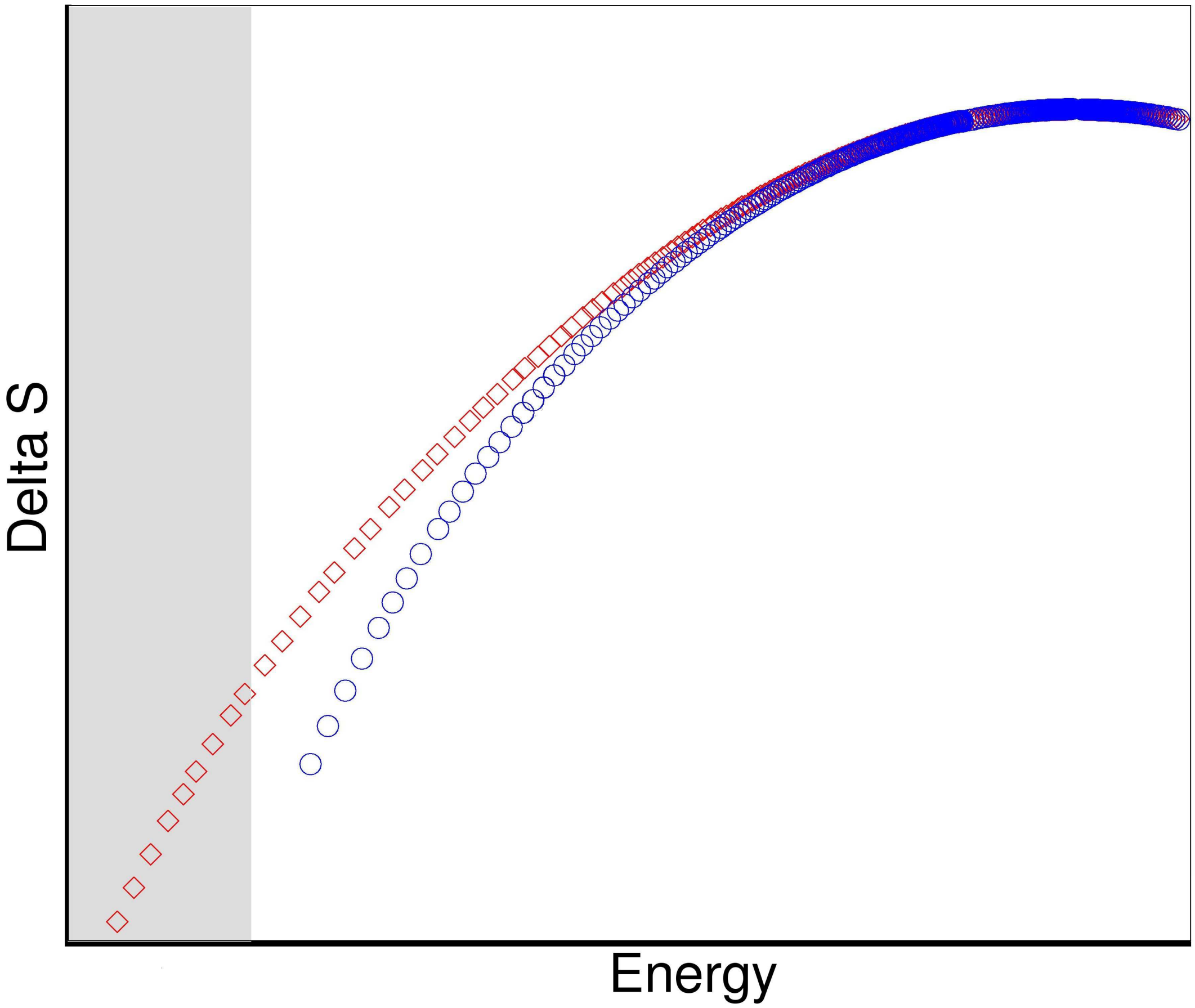

**Figure 4**